# A deterministic model of the free will phenomenon

Dr Mark J Hadley

Department of Physics, University of Warwick

## Abstract

The abstract concept of indeterministic free will is distinguished from the phenomenon of free will. Evidence for the abstract concept is examined and critically compared with various designs of automata. It is concluded that there is no evidence to support the abstract concept of indeterministic free will, it is inconceivable that a test could be constructed to distinguish an indeterministic agent from a complicated automaton. Testing the free will of an alien visitor is introduced to separate prejudices about who has free will from objective experiments. The phenomenon of free will is modelled with a deterministic decision making agent. The agent values 'independence' and satisfies a desire for independence by responding to 'challenges'. When the agent generates challenges internally it will establish a record of being able to do otherwise. In principle a computer could be built with a free will property. The model also explains false attributions of free will (superstitions).

**Key words**: free will; Determinism; Quantum theory; Predictability; Choice; Automata;

A deterministic model of the free will phenomenon

# 1. Introduction

We challenge the evidence for indeterminism and develop a deterministic model of our decision making which makes new predictions.

The relation between free will and physics is contentious and puzzling at all levels. Philosophers have debated how free will can be explained with current scientific theories. There is debate about the meaning of the term free will, even leading to questions about whether or not we have anything called free will. A key focus of the philosophical debate is compatibility of free will with deterministic physical theories. Philosophers who argue against determinism, suggest a fundamental role for quantum theory in models of our decision making. It is the supposed link to quantum theory first attracted my interest. The literature extends from philosophy journals to science publications (Conway and Kochen 2006, Libet 1985, Nichols 2011).

This work takes a unique approach to the problem, looking for evidence, building models and making predictions. It is critically important to recognise two different uses of the term free will. An abstract concept, and a known property of human decision making, they are distinct and require different approaches, but they are often confused. Searle (2007) points to the lack of progress on the free will problem over centuries and suggests that the way forward will be to recognise a false supposition. We identify that false supposition that: the phenomenon of free will provides evidence and relevance for the abstract concept of indeterministic free will. It does not.

There is an abstract concept of indeterministic free will. It is the concept of a decision making process not governed by classical deterministic laws of physics. Because this is an abstract concept, it makes sense to ask '*Do we have free will?*' If we understand the concept then we can design tests to answer the all-important question '*Do we have free will?*' The answer might be expected to depend on exactly how we define the conceptual form of free will. For the abstract concept called free will we ask what its properties would be and how we could test for its existence or measure it.

This paper also recognises a phenomenon of free will that we possess as a characteristic of human decision making - a belief and common experience that *we could do otherwise*. It is widely accepted, almost universal, and crosses cultural divides (Sarkissian et al 2010). It underpins theological, legal and moral systems (Nahmias et al 2007), (Nichols and Knobe 2007). The overwhelming majority of philosophers and commentators ascribe the property to humans, generally not to animals, and most definitely not to computers. We will try to characterise and model the phenomenon and then test the model against the facts. Note that the phenomenon of free will (the phenomenon) exists, it is up to us to accurately model the phenomenon. We will do exactly that.

This is not a review paper. Philosophical and other references are given to respected sources to illustrate the debate, rather than as a comprehensive review. This paper is exclusively about the decision making process. Some debate is about the ability or otherwise to enact a decision, where an agent freely makes a decision but is impaired from acting on it by one form or other of constraint (Frankfurt 1969). What happens after a decision is reached seems relatively free from paradoxes and does not challenge the interface between the mind and the laws of physics.

In the literature the same term, free will, is used for the abstract concept of indeterministic decision making and also for the phenomenon that we can do otherwise, which is a cause of substantial confusion and is at the heart of most assertions that quantum theory is required to explain free will. Some authors recognise the assumption they are making (Searle 2007), others seem to make it unwittingly.

A deterministic model of the free will phenomenon

Arguments along the lines of: free will [the concept] is incompatible with deterministic laws; we have free will [the phenomenon] therefore it must be due to non-deterministic theories, of which quantum theory is our prime example. Confusing the two also takes away any motivation to look for evidence of the concept, because the phenomenon is taken as that evidence. The confusion also undermines the search for models because decision making that is indeterministic is equated to free will (the concept) without explaining why that gives rise to perceived *freedom to do otherwise*, which is the phenomenon of free will.

For the clearest view of the conflict between free will and scientific theories, we look back to Victorian times. At the start of the twentieth century laws of physics were known and well tested. These were classical laws like Newtonian mechanics and gravitation plus statistical laws like thermodynamics, it looked to many people as if science was close to a full and complete description of Nature. The laws were deterministic: once you knew the initial conditions what happened next was predetermined. Even if one did not actually know the initial conditions, it was presumed they existed and the deterministic laws of physics applied and determined subsequent evolution. Probability distributions could be used to put a measure on our ignorance of those initial conditions. Apparent randomness, as in a coin toss, was just an artefact of our limited knowledge of initial conditions. The gas laws, for example, were derived from the motion of molecules. Average properties were accurately described even though individual molecular trajectories could never be measured in practice.

The Victorian era also gave us increasingly sophisticated automatons in shop windows and fair grounds. Some played music and were programmable e.g. with interchangeable discs to control the notes and play different tunes. The operation and behaviour of the automatons was clearly compatible with, and described by, the deterministic laws of classical physics. They might commonly be described as clockwork models, in principle today's computer controlled robots would also be classed as automatons. The abstract concept of an agent having indeterministic free will is the antithesis of being an automaton.

Today we know that the world is ultimately described by quantum theory. It is theoretically impossible to define precise initial states (note that it was always practically impossible to do so) and outcomes of experiments are intrinsically indeterministic. The atoms in our bodies can only be described with quantum theory. In the free will debate the question is not whether indeterministic laws of physics apply to us – they do, but whether or not we need to invoke quantum indeterminacy to accurately describe our decision making. For example an automaton, made of atoms, can be described adequately with classical, deterministic laws of physics (or engineering).

The first part of the paper looks for evidence of the abstract concept associated with free will – indeterministic decision making. We design tests and look at incorporation of randomness and quantum theory with an emphasis on experimental tests. In the process, an increasingly complicated automaton is described to show how simple tests are easily misled. The types of randomness and the relevance of randomness is illustrated with examples to clarify and challenge claims that it is a key feature of free will.  A definitive test of tests is introduced.

The second part builds a model to explain the phenomenon of free will. This is set against known tests and also makes new predictions. While the many philosophers are compatibilists, believing that the phenomenon of free will is compatible with deterministic laws of physics (Dennett 1984, Smilansky 2000), models of such decision making have been unconvincing and untested.

## 2. Neurology and psychology

What do we know about our own individual decision making? If we use introspection to assert as a fact that we have free will that implies recognising some feature of our decision making that is free. Far more fundamentally it assumes that we are aware of our own decision making. There have been some ground breaking studies of the brain during decision making that raises big questions about when decisions are made in our brain and our awareness or control of them.

In Libet's famous experiments (Libet 1985) subjects made decisions and pressed a button when they had reached a decision. At the same time brain activity was monitored, which seemed to show activity associated with choice which occurred before the subject was aware of making a decision. There has been intense debate about how convincing the experiments are and alternative ways to interpret them. They are certainly fascinating experiments that will continue to be repeated and refined. For our purposes, the most important thing it tells us is already well known and supported by experiments in psychology. Psychologists have known for a long time that there are subconscious influences on our decision making (see for example Double 1990) Subliminal advertising is a well-known example and is now banned. Even more impressive are stage shows like Derren Brown (Brown 2006) where contestants make apparently free choices which were in fact controlled or predictable. For some acts he has publicly shown how subtle tricks influence behaviour. These are such powerful effects that he can build a career of reliable stage shows using them. The psychology literature is extensive, see for example Stanovich 1986 and the wealth of references therein.

In addition a whole range of psychology experiments repeatedly show how our decisions are subject to unconscious bias (Nisbett et al 1980, Stanovich 1986) and bookshops are full of management and child psychology books which explain how to influence the behaviour of others. While some elements are logical and transparent others are more subtle and work at the subconscious level. Subjects can subsequently add justifications and explanations for their decisions, these are after the fact and need not be accurate descriptors of the decision making process.

Libet's work raised the question *'Are we aware of our own decision making?'* the answer is either 'no' or 'maybe not'. That is sufficient for us in this analysis. *We are not necessarily aware of our own decision making process*. This completely undermines introspection as a reliable evidence of the abstract from of free will. We cannot use personal experience to declare that our decisions are non-deterministic if we cannot be sure how they are made. We cannot claim that our decisions contradict classical physics based on introspection.

## 3. Testing an automaton

To find experimental evidence for truly indeterministic free will, we look for experimental tests that can distinguish an automaton from an agent that has free will in the abstract sense. We will start by describing increasingly complicated automata. In all cases they follow the laws of deterministic classical physics and the actions can be predicted in advance by an engineer with sufficient knowledge of the workings. In principle there is no difference between a clockwork automaton with wheels and cogs and a modern electronic computer. Note that predictability, randomness and indeterminism are different concepts, systems can be built with any combination: several combinations are created in the models that follow. By definition, in a deterministic system, if we have complete knowledge of initial conditions (in practice a small subset of the full initial conditions is sufficient information) and know the rules governing evolution of the system, then we can predict outcomes. We will start with automata for which an engineer has such information, but an observer is generally ignorant of the details.

A deterministic model of the free will phenomenon

### Randomness, quantum theory and predictability of an automaton

Randomness, predictability, spontaneity and quantum indeterminism have all been related to the free will debate. The assertions persist even though there are well argued cases that they are not relevant, or even that free will is contrary to randomised decision making (Dennett 1978). Of course, by definition, the abstract concept of free will requires some departure from deterministic decision making. Crucially, a test for the existence of the concept of indeterministic free will, requires evidence to show that it is not deterministic.

In this section, we hope to clarify the different implementations of randomness and the relations between randomness and predictability. As an aid to visualisation we will start with a clockwork automaton, as our *agent*, and add features to it. Consider a fairground automaton that accepts your coin, its arms move to pour out a cup of tea, maybe adding sugar and milk and stirring it. Then finally it makes a *decision* to drink or not to drink and either leaves the cup on the table or lifts it to its lips. This would take some skill to make, but is certainly possible and models of similar complexity are in museums around the world. Our example could clearly be implemented with graphics on a computer screen or with a programmable robot.

### Deterministic and predictable

In its simplest form our agent decides to take a drink each time a coin is inserted. That is predictable and is a simple deterministic motion controlled by the wheels, cogs and levers.

We could add a level of complication. It takes a drink alternate times. That is also straightforward to create, it is deterministic and predictable. Or at least it is predictable, given the information about the construction. Without knowing the rule, it might take a little while to observe repeated cycles and form a hypothesis about the operation, leading to predictions. Note that the decision now depends upon both the external trigger of inserting the coin and on the internal state of the cogs which will vary each time: in this case alternating between two states.

We could go further so that it takes a drink every other time, or every seventh time, but not multiples of fourteen. Again, not too difficult to do with cogs and wheels. Let's implement it with a black box having a yes/no lever. If the lever is up it drinks, when down it does not drink. Inside the box is a wheel with a cog that moves the lever each go, from up to down or vice versa. There is also a wheel and a gear that goes round at one seventh the speed and similarly moves the lever from up to down or vice versa, but now every seventh attempt. Given this knowledge of the interior workings of the box, and either the past history or visibility of the cogs, the agent is deterministic and predictable. Without the knowledge of what is in the black box, the deterministic machine is not predictable. Indeed you need to know both the internal design and the internal state of the black box to make a prediction.

### Deterministic and unpredictable

The previous example had just two cogs in the black box, and predictability practically vanished. In a few minutes a computer programmer could create a deterministic algorithm with twenty or thirty factors. Similarly, the agent was triggered by a single external action: the insertion of a coin. It could easily depend on the approach of a person, how light or dark it is etc. The internal state had just two settings which could be determined from knowledge of the last seven decisions. Again this could be easily expanded to a complicated internal state with many independent factors. Just ten cogs each with ten different gear ratios and the black box would have ten billion different internal states. We could have one cog that turned very slowly changing the lever position once every 999,983 turns or higher, the agent would then appear close to spontaneous.

A deterministic model of the free will phenomenon

Note how the unpredictability comes from ignorance. The automaton is deterministic. Predictability can be restored but requires knowledge of both the design and internal structure inside the black box.

Unpredictability is a common feature of organisms with recognised evolutionary advantages. Diverse behaviour is exhibited by genetically identical samples in closely controlled environmental conditions. This has been claimed as indeterministic and used as a basis for models of free will (see for example Brembs 2011). As can be seen above deterministic systems can be unpredictable, for all practical purposes, as their complexity increases.

### Random and unpredictable

We now seek to add randomness to the workings of the black box. How to generate random numbers is a substantial academic topic in its own right, here we will give a few diverse examples.

A traditional way to get a random number was to use a table of random numbers. Our agent could have such a list built in to it and look up the next number on the list and act according to the number being odd or even. That would be trivial to implement in a computer program. For our mechanical agent, this can be implemented using a cog wheel where cogs were missing or present according to list of random numbers – much like the discs in a musical automaton. That is a strange implementation, the decisions are now clearly predetermined, but have all the statistical characteristics of random numbers.

Another approach is for the agent to use an environmental variable to generate a random number. A simple example is to incorporate a fine clock in the black box that counts in microseconds or nanoseconds. When the coin is inserted the number of nanoseconds is used to set the lever. In the simplest case depending upon the number being odd or even. The environmental number could even be used to select one of several random number tables (the term seed is used for similar systems).

We have described an automaton with an internal source of randomness. It is still an automaton with outcomes described by deterministic laws of physics from an initial state. This is a type of model that appears in the philosophy literature as a 'two stage model' where randomness of one sort or another is a seed or influence at an early stage followed by a deterministic, rational choice process.

### Quantum uncertainty and predictability

The examples above all use classical physics. There are reasons to believe that the randomness in quantum systems is fundamentally different to classical randomness. The latter is based on us lacking knowledge of the initial conditions. There are powerful theorems, supported by experimental evidence, that quantum probabilities cannot arise from unknown initial conditions. The term used in the literature on foundations of quantum theory is hidden variables – quantum theory is incompatible with any local hidden variable theories.

The simplest of quantum systems might use spin properties of particles to generate an indeterministic 50:50 decision. Such a quantum based decision making system would be random. But if it operated inside a sealed black box it would be indistinguishable from a system using random numbers or pseudo random numbers. In turn all these would be practically indistinguishable from a black box containing a complex arrangement of wheels and cogs, provided that the mechanical complexity was large compared with the number of decisions being analysed.

As a technical note: there are some experiments that distinguish quantum randomness from any possible classical system. Such tests use pairs of entangled particles with large physical separations. The systems need to be isolated from the environment using high vacuum and very low temperatures. Even then the distinguishing features only arise by looking at probability distributions from a large number of instances. It is inconceivable that the conditions exist in our heads, and there is absolutely no evidence that they are a feature of our decision making.

### *Conclusion*

It is easy to envisage a clockwork agent that appears to act unpredictably. Without a knowledge of the construction, without seeing inside the black box, it is impossible to distinguish randomness from pseudo randomness or from a complex clockwork arrangement (Dennett 1984). Quantum theory offers no discernible difference in behaviour compared with decisions that are classically random or pseudo random.

All the examples above, excepting true randomness and quantum randomness, are such that replicas could be made and if we exactly copied all the internal structure and set them up identically, and the coin was inserted at the same exact nanosecond, they would all make identical decisions. That is not free will. But we also know that for an individual system, randomness in the decision making is externally indistinguishable from pseudo randomness, even though in principle the latter can be replicated. This is simply seen by using a classical or quantum random number generator recording the numbers and then encoding them on to gear wheels.

## 4. Evidence for the abstract concept of indeterministic free will

The abstract concept of indeterministic free will, is very technical and has a precise meaning in mathematics, physics and philosophy. The language may be deceptively similar to descriptions of the phenomenon of free will, but the latter is more of a folk tradition than a statement about theoretical physics or quantum theory. To investigate the relevance of an abstract concept we ask what the evidence is and what tests could be used to search for it. We make two claims:

**Claim 1.** *There is no evidence for indeterministic free will.*

**Claim 2.** *Furthermore, there is no conceivable test to distinguish the decisions of a deterministic agent from an agent making indeterministic decisions.*

Although the claims have been motivated by considering a very simple automaton like decision maker, they will be confirmed later with more realistic models. We are certainly very complicated agents. Our own intuition and introspection are known to be unreliable. In a wide range of situations we don't know how we reached particular decisions. The claim that our subconscious awareness of decision making is post factual cannot be refuted. Therefore introspection does not provide any reliable or credible evidence for the concept of indeterministic free will. Discussion of the automaton with increasing levels of complexity, shows that an analysis of the decisions that an agent makes will not be able to provide any evidence for indeterminism.

There is no evidence. There is no conceivable experiment or test that distinguishes human decision making from that of a complicated automaton following the deterministic laws of classical physics. We believe that the concept of indeterministic free will is irrelevant to an understanding of the phenomenon of free will.

While this dismissal of indeterministic free will might seem extreme, it is not dissimilar to the views of leading philosophers (such as Dennett 1078 and Smilansky 2000), who argue that we don't have indeterministic free will as defined by the abstract definition. They argue instead that we have some flexibility in our decision making that gives us the illusion of freedom. Dennett even proposes a model where some type of randomness affects the number of different factors we use when making decisions. Although his model is unconvincing, the argument is that our experience of decision making and perception of free will may be accountable within the laws of classical physics. In short that a phenomenon of free will is compatible with determinism.

## 5. Tests for free will

In the next section we aim to build a model of free will. As scientists we want to test the model and even make predictions. It is our belief that the lack of objective tests has hampered the study of free will and allowed unsatisfactory model proposals to persist. However there is an overwhelming prejudice that we have to overcome in order to develop objective tests. We associate humans with having free will and we assume that familiar metal objects, including computers, do not have free will. To this end we will consider an agent from outer space.

### An alien agent

Consider that a rocket lands on Earth. The doors open and out comes a figure in a sort of spacesuit. Let's imagine a humanoid looking figure. From its appearance we can't tell if it is an intelligent agent like us (As much a free agent as us) wearing a spacesuit, or an automaton, programmed by an advanced alien race. How do we decide? What tests can we apply?

Imagine that the alien walks around and sets up experiments? Does that help? Suppose it can communicate, either it learns our language or already speaks one of our languages. Does that help? Could we ask questions that would determine if it were a free agent?

We could ask it if it had free will and might give some credence if it said *'Yes'* But such a response would be trivial to incorporate into a computer controlled robot.

What if we dismembered and dissected it? Could that lead to a test?

### Test of tests

If we can devise a test for the alien then we can apply similar tests to humans, to dolphins and other animals and even to sophisticated computer based systems, not to mention our clockwork automata. A positive outcome for humans is essential for a credible test of the phenomenon of free will. We will then be a long way towards a model of the phenomenon of free will.

Our test of tests, is that a candidate for a test can be applied to the alien agent in such a way as to decide if it is an automaton or not. We think it is critical to have such a test and we propose one below. The alternative is little more than prejudice when we declare that the automaton at the fairground does not have free will, but that the pickpocket does!

To reiterate our claim that there is no evidence for the abstract concept of indeterministic free will. The alien agent puts that claim into perspective. We challenge anyone defending the abstract concept of indeterministic free will to explain `*How would you test the alien agent for it?*'

At the other extreme, unpredictability is no test of free will, because that is readily provided by an automaton. In deciding that the automaton models with alternative actions do not have free will and that a Geiger counter with indeterministic actions does not have free will, we are subconsciously applying a test. To proceed with a model, as scientists we need a test.

A deterministic model of the free will phenomenon

### A challenge test for free will

We construct the following test, that we think encapsulates our perception of our own free will and also how we recognise it in others. It is a test for the phenomenon of free will, not for an abstract concept like mathematical indeterminism.

**Definition: Exceptional action** *is a rare action of no apparent value to the agent. It is unlikely, unfavourable, or even has significant adverse consequences.*

**Definition: Highly discerning test** *looks for evidence of free will by prompting an exceptional action.*

The idea is that an agent can demonstrate that it *could do otherwise* by making a decision to take an exceptional action. By inference we conclude that if it has the freedom to do otherwise for an exceptional action then it probably has the same freedom on other actions. Many actions are unpredictable, so that we get no information from one choice compared with another. However taking an exceptional action is rare and otherwise inexplicable.

An example of an exceptional action could arise choosing which hand to hold a pen with. A hundred or a thousand times we will use the same hand, but if asked we will say that we *could have done* otherwise. We can choose to use the other hand as an exceptional action to demonstrate that we could do otherwise. More dramatic examples would be to put a hand near a flame or into icy water.

**Definition: Free will test** *We challenge an agent to take an exceptional action. If the challenge results in a change of behaviour then we conclude that the agent could do otherwise.*

The free will test is intrinsically statistical, but the nature of exceptional actions is such that a conclusion could be reached after a few repetitions.

This leads to a test for the alien agent. Not a test for the abstract concept of indeterministic free will, but for the phenomenon of free will that we recognise through experience of our decision making. We would challenge it to do something and see if the challenge altered its behaviour. To be a highly discerning test we should find an action that would be highly unlikely otherwise, maybe one that is risky or moderately harmful. For example, we could challenge it to approach a fire. It might express logical reasons not to. The test is to challenge it to show it has freedom by getting close to a flame anyway. One request in one scenario would not prove anything given the complexities of the environment and the alien, but if the alien always made logical decisions independent of our challenges, then it would be perceived as an automaton. Alternatively, if it responded and took unprecedented actions in response to challenges, then it would appear to have free will.

> **Claim 3.** *Alien test: We would see if an agent had free will (the phenomenon) by challenging it to do an exceptional action. If it responded to the challenge, the test would be positive.*

## 6. The phenomenon of free will

We have dismissed the abstract concept of indeterministic free will as being unsupported by any evidence. The alternative scientific explanation is that a deterministic model based on classical physics could be constructed. In philosophy terms we are compatibilists. We are in good company. But we do not actually have a model.

There is a widespread, almost universal, cross cultural belief that our decision making has a property called free will: that we *could have done otherwise* (Sarkissian et al 2010). Such a widespread perception deserves attention and needs an explanation. It is our challenge to us to try and model the phenomenon of free will, test it against real life and make predictions. To be precise we want to model our perception of having free will. It is quite a reasonable expectation, there are many aspects of our perception that are technically false but can still be explained - for example why metals feel colder than plastics even when they are at the same temperature.

We want a model that will pass our free will test.

## 7. The inadequacy of indeterministic two stage models

Current models in the literature are predominantly two stage models, often with some randomness (indeterminacy of one sort or another) involved at an early stage, generally subconscious, followed by a rational (or more precisely conscious) choice between the possibilities in the first stage. Dennett (1978) dismisses true indeterminism, but incorporates a weaker version into the factors considered for a decision, Kane (1985) wants randomness in the final decision making. Long and Sedley (1987) talk of atoms swerving in unphysical ways. Kosslyn (in Libet 2009) describes models with indeterminacy based on chaos theory. As Searle says, the models are unconvincing. Many models are constructed to implement the abstract concept of free will where choices are not deterministic. They are not tested against experiment, they don't make predictions, and they don't easily apply to *highly discerning tests*. They do successfully describe a decision making agent that is not deterministic, despite the lack of evidence for such an agent. The lack of evidence means that there is consequently nothing to test the models against - they are irrelevant.

The models are clearly not necessary because any pattern of decision making or internal workings of the mind due to quantum indeterminism can be replicated by a deterministic process as described for the automaton earlier.

Neither are they sufficient to explain or recreate free will. Strangely, the models are so simple that it would be easy to make a computer with their model of free will. We can, for example, connect a process like radioactive decay to a rational processor. The radioactive decay is an example of quantum indeterminacy. Using that as one input to a simple processor, is a two stage model of action. It gives a meter reading or robot arm moving that depends in some way on the incident radiation. It is a variation of a Geiger counter. It has all the elements of a two stage model, but it has never been suggested that a Geiger counter has free will. Advocates of the two stage model have never taken the logical step of creating free will in a machine to prove their models.

> **Claim 4.** *Two stage models of free will based on quantum indeterminacy are neither necessary nor sufficient to explain free will.*

## 8. A model of free will

Let us start with a complicated but logical decision making model. Generic models of agents are classified by Russell and Norvig (2013). In particular we will use model-based, utility-based agent, a goal based decision making which includes internal measures – the utilities. Degrees of happiness is often given as an example of a utility. This type of agent is used widely across many disciplines. In economics, sociology and psychology it is used to model aspects of human decision making; in engineering it is used in models of control systems and in computer science it is a basis for autonomous, intelligent systems. It applies to a person making a purchasing decision, a drone avoiding obstacles or a mobile phone conserving battery power. It makes no presumption of intelligence, consciousness or even if it is organic biological or inorganic: it is a very general model. The agent has a decision to make, and several, possibly competing goals, with different weights. There are several environmental inputs.

A deterministic model of the free will phenomenon

The agent is capable of logical analysis which need not be perfect, it is model-based in that it has a model, not necessarily perfect, to predict the effect of its decisions. Additionally we include some inputs from internal states of the agent- the utilities.

Let's have an example to illustrate. Will the agent have a second biscuit with their drink? Picking a second biscuit or leaving it on the tray is the decision. Competing goals might be satisfying hunger, satisfying a sugar craving, social factors like maintaining social respect, wanting to enjoy a meal out later on, wanting to loose (or gain) weight. Environmental inputs would be the look of the biscuit, what other people were eating, what the time was on the clock etc. A logical analysis might be that it is two hours until dinner and the effect of the biscuit on appetite. Another logical analysis will be some prediction of how enjoyable the second biscuit will be, informed by the memory of the first one. Examples of internal states might be how hungry the agent is, memories of eating the first biscuit, or how socially comfortable the agent is.

Such a decision making scenario is not trivial, and would be inconceivable to implement with gears and cogs, but quite feasible on a computer system. We could implement it with an algorithm programmed into a computer. It is deterministic. It could be highly predictable, but an element of unpredictability can easily be added. A simple approach is for the algorithm to give an output between 0 and 1, with 0.5 and over being the threshold for taking the biscuit. Alternatively the threshold could be a random number between 0 and 1 rather than simply 0.5. Such a method gives some variability but preserves the integrity of the decision making process. Dennett suggests that the factors can vary to some extent randomly which is another way to add unpredictability. A similar effect would be achieved in a completely deterministic agent by having a longer list of weightings and dependence on other environmental variables and internal states, some of which were hidden.

Although not part of the decision making, it may help an observer relate the actions to the free will question if the agent were able to explain the decision making. Again this is not difficult, either as a numerical list of factors and weights or encapsulated in words like 'I was really hungry but I did not want to seem greedy and on balance I chose to....' People do this and it is quite feasible to implement in a computer system.

Does the agent have free will? Does the variability from the randomness manifest itself as free will? If the decision making is repeated many times, an outside observer might see an element of unpredictability, but no more so than from an automaton with an unknown mechanism. The agent themselves would be able to report a string of decisions with some variability, indeed if the weightings and inputs had a strong bias then the variability would be minimal and the agent could reasonably predict that the next decision would be no different. Freedom would be no more or less than control over a reflex like hiccups or heart rate.

An outside observer would see alternative possible outcomes and be unable to predict which will occur. The agent could do a range of different actions, but no more so than even a simple automaton described earlier. We have implemented unpredictability. But unpredictability is commonplace, the free will phenomenon needs more than unpredictability to describe it.

In our simple model the question *'could you do otherwise?'* has no effect, it does not affect any of the defined goals, and the question is not an input to the algorithm. We seek to implement free will in the simplest most direct way. There may be other ways to achieve a similar outcome. Indeed we hope other authors will develop better models that can be tested and compared with experiment.

A deterministic model of the free will phenomenon

We proceed by explicitly adding an extra factor. We define an extra utility goal, call it *independence* (curiosity might be an alternative term) which is satisfied by responding to an external *challenge*. Of course it presupposes that the agent is sophisticated enough to recognise a challenge.

> **Definition: Independent *agent*** is a decision making agent that has a property called *independence as one of its goals. It satisfies the goal by responding to a* challenge 'to do otherwise'. *The agent necessarily has the capability to recognise challenges.*

Now we can challenge a greedy agent *'Did you have to take the biscuit?'* and the weighted factors swing away from taking a biscuit to give greater satisfaction by leaving the biscuit. With communication, the agent can explain its actions *'I was hungry, the biscuit was tasty, but I did not have to take it'.* However we must stress that such an explanation only serves to help an observer analyse the agent, it is an independent agent, with or without the explanation.

An exceptional action illustrates the power of this one extra goal and input. Consider an agent that has to pick up a pen and write a signature. Almost without exception our agent uses its right hand. Factors like efficiency or custom and practice will weigh so strongly that the right hand is always used. The picture is dramatically changed when we add the independence goal and give challenges to the agent. Now in response to a challenge, the agent will commonly pick up and write with the left hand. The agent can explain its actions by saying '*Using my right hand is much better but I can choose otherwise*'. Even in a fully deterministic system we have created an agent that *'Could do otherwise'*.

To summarise, we have a deterministic model of free will. We have a model based utility based decision making agent responding to inputs and satisfying several competing goals. All these are common in modern computers. We then add one extra goal called *Independence* which is satisfied by responding to a *challenge*. Challenges can be external or arise internally to the agent. We call this an independent agent. In principle the independence goal is no different to others like being well-fed or warm, though it is more at an abstract or emotional level like wanting to be liked, or exercise creativity.

> **Claim 5.** *We have defined the structure of an agent that 'could do otherwise'. It includes a goal of* independence *and satisfies that goal by responding to* challenges.

> **Claim 6.** *Our model shows conclusively that the phenomenon of free will is compatible with determinism.*

## 9. Do we have free will?

Yes. There is a phenomenon which we call free will, a perception that we could do otherwise. We have modelled it accurately. The model is deterministic in a mathematical sense. However the model shows that in common language, we can do otherwise, we know we can do otherwise and others see that we can do otherwise. Nothing is missing from the common sense notion of free will.

## 10. Why did it take so long to understand?

As John Searle (2007) said "*The persistence of the traditional free will problem in philosophy seems to me something of a scandal. After all these centuries of writing about free will it does not seem to me that we have made much progress*". He goes on to predict , "*when we at last overcome one of these intractable problems it often happens that we do so by showing that we had made a false presupposition*" By focusing on evidence and models, this work identifies two shortcomings in the traditional debate:

The folk intuition about free will uses the same words and phrases as used to describe indeterminism in mathematical physics (and philosophy): "there is more than one alternative action" "the future is not predetermined." This is unfortunate and terribly misleading. The folk intuition was never talking about evolution from a fully determined set of initial conditions. It never could have been. Our minds simply do not have access to that microscopic level of detail. Our observations and experience do not have the precision to meet the mathematical requirements for repeatability. The mathematical concept of determinism is a very precise, abstract concept, primarily applicable to simple physical models. Academics have mistakenly linked free will to mathematical indeterminism. In doing so they created a problem that could not be solved. Mathematical indeterminism is the false presupposition that Searle refers to. We are not just dismissing it as a solution, we claim it has no place in the free will debate and never should have.

The second, related, source of confusion was the degree to which introspection was taken as a guide to reality rather than simply our perception of reality. We don't know or understand how we make decisions. We think "we can do otherwise" but that is not evidence for mathematical indeterminism. This is the false presupposition. When someone is knows thst they could have done otherwise. Is this a statement that they know that if every atom was in the same position, momentum etc., and every detail of the environment was reproduced with microscopic precision, that the outcome could be different? Of course not. People have such knowledge and generally don't express free will at that level. We have dismissed such a suggestion as unfounded. Or do people mean that on a different day, or in a different mood, or if prompted differently, or self reflected first, then the outcome might be different? They do mean that, and it us exactly what we have modelled.

We have a perception of indeterministic decision making, it is the perception that needed to be explained. Too much of the debate is about mathematical indeterminism which is not relevant, and proposing models that don't explain our perception.

## 11. Enhancements to the agent model

The model described above implements free will in a deterministic system. It is relatively simple and crude. Some obvious, but optional, enhancements take the model closer to the human experience and have some explanatory power.

### Communication

The agent needs some level of communication in order to recognise a challenge. The communication need not be linguistic and can be one way. The agent simply needs to recognise a challenge in its external environment. It would however be most helpful if the agent could convey its reasoning process with an explanation. E.g. "It is painful [damaging] to go near the flame, but I can do it" or "I avoided the flame because it is dangerous, but I could do otherwise"

Note that in the examples the claim to do otherwise is matched by the pattern of decisions and actions. It is not simply a programmed verbal response.

### Self reflection and analysis

The agent could record a history of actions and draw conclusions, such as "I usually do this, but when asked if I can do otherwise." Such a capability is common in systems related to databases.

> **Claim 7.** *An independent agent that can analyse and report its actions will report being able to do otherwise.*

### Self generated challenges

The ability to generate challenges internally is very powerful, particularly combined with some analysis and reporting. This can be a purely logical deterministic process, a type of curiosity perhaps driven by other goals known or unknown. The agent can then ask itself *'Can I do otherwise?'* It can add the challenge to a decision that would otherwise be predictable. It is a rich and complicated feedback process capable of being implemented at several levels of abstraction. Note that the process is pathological if the agent knows its own construction.

### Handling abstract concepts

Some computer based systems can manipulate abstract concepts – Mathematica handling algebraic expressions is a well known example. If an agent can process a concept of free will, it can relate its record of decision making to the concept. A higher level would be to generate its own challenges. That might be discernible to an outside observer. Imagine the alien agent testing a flame, withdrawing from it sharply, then trying again. It could test itself with "could I do otherwise" and conclude that it could. Given the abstract concept of indeterminism, it could test itself and conclude that its actions were not predetermined.

> **Claim 8**. *If we cannot communicate with an agent then we will have difficulty ascertaining freedom of action, but we may recognise signs of responses to self-generated challenges.*

Note that such an independent agent will *know* that it *could do otherwise*. It will *know* that any apparently predetermined or predictable action can be interrupted and stopped. It can test that hypothesis by generating a challenge and noting the change. It will conclude that its actions are not predetermined. The whole process just described can be entirely deterministic, but the experience will not be.

> **Claim 9**. *An independent agent that lacks full knowledge of its decision, but is capable of abstract thought, will develop the concept of free will to describe its decision making.*

## 12. Predictions and further work

The abstract concept of free will: decision making that is not a deterministic consequence of the laws of physics, has been dismissed for lack of evidence. Our knowledge and awareness of our own decision making is unreliable and incomplete. There is no experimental evidence for a lack of determinism in our decision making and there is no conceivable test that prove otherwise.

The phenomena of free will, that when we make a decision, *we could have done otherwise* is modelled with a deterministic algorithm. The agent has a goal of independence (amongst many others) which is satisfied when it responds to a challenge. The agent is sufficiently complicated to be able to recognise a challenge: we call this an independent agent.

> **Claim 10**. *An independent agent will be perceived as having free will*

> **Claim 11**. *An agent without the independence property will be perceived not to have free will.*

> **Claim 12**. *Free will is not an illusion. The agent can do otherwise when challenged or when it generates its own challenge.*

Our model also explains false attributions of free will. Twentieth century high technology civilisations recognise natural phenomena such as the weather, volcanoes, tides etc. as forms of fluid flow, governed by complicated equations of fluid mechanics and thermodynamics. In practice they are unpredictable because we don't know the initial conditions and we don't have the computing power. In other cultures the systems are given personalities and god-like status. They are attributed free choice, and the ability to respond to human behaviour through prayer and sacrifices etc. Unrelated to the free will issue, our brain has a remarkable inclination to look for and find patterns in events (Ebert and Wegner 2011, Brown 2006), even in random events where it is a characteristic known as apophenia or patternicity even when they are actually random; there are evolutionary reasons why this should be so (Foster and Kokko 2009, Langer 1975). While our culture would say that weather and other natural phenomena are unrelated to our petitions and offerings, other communities might perceive a correlation. If they see a request to the gods to abate the weather followed by an improvement, they will assign free actions to the gods. They are in effect doing the alien challenge test, perceiving a correlation which reinforces a belief that the weather or a god or volcano has free will and can choose what it does.

> **Claim 13.** *False attributions of free will are made by applying the alien test and mistakenly perceiving a correlation. This gives rise to some superstitions.*

The moral implications of this deterministic model of free will are not as severe as one might have expected. We could add to the agent an ability to learn from experience and modify the weights attached to different goals. This can still be deterministic and is readily achieved with today's programmable computers. Most concepts of crime and punishment are still valid. Of course this is a necessary consequence of having a good model of free will. Our aim was to model the phenomenon of free will, we have done so with a system that responds to challenges, and challenges its own decision making. It is a good starting point for theories of reward and punishment.

> **Claim 14.** W*e could build a computer system with free will. This follows directly from the independent agent that we have described.*

It is likely to be a highly contentious claim because we have a long established prejudice that humans have free will and mechanical objects do not.

To do further work and test algorithms such as the independent agent, we would recommend the use of cartoons, where form and behaviour are independent and under the control of the animator. For example the robot, Bender, in Futurama looks like a tin can but behaves with the characteristics of a deviant human. Conversely Spock on Star Trek is portrayed as being entirely logical (not always convincing), but takes human form. Our prediction is that implementing the independent agent algorithm in cartoons, will give the audience the perception that the character has free will. And conversely that if the character shows no signs of changing behaviour in response to challenges then the audience will not attribute free will provided that the visual appearance is neutral. The audience can answer the question *'Could they have done otherwise?'* The medium of cartoons can be used to test related theories and other algorithms.

As one important element in a model of human decision making, the independence factor and the response to challenges can help develop better models of human behaviour with the potential to inform addiction strategies.

No doubt the model can be refined and possibly even challenged. But we invite responses that are evidence based and alternative models should be testable, distinguishable and refutable.

A deterministic model of the free will phenomenon

A deterministic model of the free will phenomenon